\documentclass{elsart}

\usepackage{natbib}

\def\simgt{\mathrel{\raise0.35ex\hbox{$\scriptstyle >$}\kern-0.6em
\lower0.40ex\hbox{{$\scriptstyle \sim$}}}}
\def\simlt{\mathrel{\raise0.35ex\hbox{$\scriptstyle <$}\kern-0.6em
\lower0.40ex\hbox{{$\scriptstyle \sim$}}}}

\newcommand{\be}{\begin{equation}}
\newcommand{\ee}{\end{equation}}
\newcommand{\bea}{\begin{eqnarray}}
\newcommand{\eea}{\end{eqnarray}}

\input epsf
\usepackage{amssymb}

\begin{document}

\begin{frontmatter}



\title{Finding SZ clusters in the ACBAR  maps}


\author{ E.~Pierpaoli}
\address{California Institute of Technology,
 Mail Code 130--33, Pasadena, CA, 91125~~USA}

\author{S.~Anthoine}
\address{Dep. of Applied Mathematics,
 Princeton University, Princeton, NJ, 08544~~USA}

\begin{abstract}
We present a new method for component separation from multifrequency maps
aimed to extract Sunyaev-Zeldovich (SZ) galaxy clusters from Cosmic  Microwave 
Background (CMB) experiments.
This method is best suited to recover non--Gaussian, spatially 
 localized and sparse signals.
We apply our method on simulated maps  of the  ACBAR experiment.
We find that this method  improves the reconstruction of the integrated
$y$ parameter by a factor of three with respect to the Wiener filter case.
Moreover, 
the scatter associated with the reconstruction is reduced  by 30 per cent.

\end{abstract}


\end{frontmatter}

\section{ Introduction}
\label{}



The study of the Cosmic Microwave Background (CMB)  has greatly  improved our
understanding of the Universe in the last decade.
The measurement and interpretation   of the CMB power spectrum  has allowed 
to determine the most important cosmological parameters with very high 
accuracy.
More experiments, now planned or undeway, will produce higher resolution
multi--frequency maps of the sky in the 100--400 GHz frequency range.
One of the most important new scientific goals of these experiments is the
detection of clusters through their characteristic Sunyaev-Zeldovich (SZ)
signature (Sunyaev \& Zeldovich 1980).
Because the SZ signal is substantially independent of redshift, 
SZ clusters will be observed at very large distances, quite independently from
their mass.
Such clusters  may be used to infer cosmological information
via number counts   and power spectrum
analysis  of SZ maps.
Many studies have shown the great potential of these new technique.
These estimates, however, typically assume that all clusters 
above a certain flux are  perfectly reconstructed and detected in the CMB maps.
In practice, this may not be the case.
SZ clusters  have radio intensities comparable to other intervening
cosmological signals like the CMB and point sources.
Despite the different frequency and spatial dependence of these signals,
 it is not so easy to disentangle them.
Moreover, beam smearing and instrumental noise play a role in our ability
to adequately reconstruct the observed cluster. 
These arguments raise the necessity to assess how well a certain technique 
performs in reconstructing the cluster signal given the experiment
 specifications.
For a given cluster Compton parameter $y$,
the reconstructed value may also depend on the cluster location and shape.
Therefore, there is an error  associated with  the reconstruction 
technique which needs to be assessed and  accounted for  when it 
comes to relate cluster's observables with cosmological models.
Moreover, the specific observable to use may depend on the type of experiment
in hand. 
In this paper we address these issues for the ACBAR experiment (see 
Pierpaoli et al. 2004 for a similar analysis applied to Planck and ACT).
Several techniques have been developed for image reconstruction in the 
multi--component case (Herranz et al. 2002, Stolyarov et al. 2002). 
In most cases, these techniques are optimal in reconstructing the 
CMB  fluctuation signal, which is Gaussian and well characterized 
in Fourier space.
Clusters of galaxies maps present very different features from
CMB ones, in particular: {\it i)}
clusters are ``rare'' objects in the map, they don't fill out the 
majority of the space; {\it ii)} The cluster signal is 
non--Gaussian on several scales, 
and in particular on scales associated 
with the typical core size; {\it iii)} 
Different scales are correlated in Fourier space.
Keeping these characteristics in mind, in 
this paper we develop a method aimed to better reconstruct the SZ galaxy 
cluster signal from multifrequency maps. 
Our map reconstruction 
method is wavelet based and is best suited to reconstruct  the specific 
non--Gaussian signal expected in galaxy clusters maps.
In this work we use the simulated cluster maps by Martin White available at:
{\it http://pac1.berkeley.edu/tSZ/}.
These maps are not full hydrodynamical simulations, the gas here has been 
introduced using the dark matter as a tracer. However for the kind of experiment in hand which would not resolve the cluster structure anyway, this should not present a major problem. The underlying cosmology corresponds to the concordance model. We used ten maps of 10 $\times$ 10 degrees.

\begin{table}
\caption{The characteristics of the ACBAR experiment.}
\label{tab:ic}
\begin{center}
\begin{tabular}{ccc}
 $\nu$ (GHz) & FWHM (arcmin) & noise $ (\mu {\rm K}/{\rm beam})$  \\ \hline 
\hline
 150 & 5  & 6 \\
 220  & 5  & 10  \\
 280  & 5 & 12  \\
\hline
\end{tabular}
\end{center}

\end{table}

\section{The reconstruction method} \label{par:recmet}

\subsection{The wavelet decomposition used} \label{par:wavdec}

\begin{figure}
\begin{center}
\leavevmode
\epsfxsize=2cm \epsfysize=2cm \epsfbox{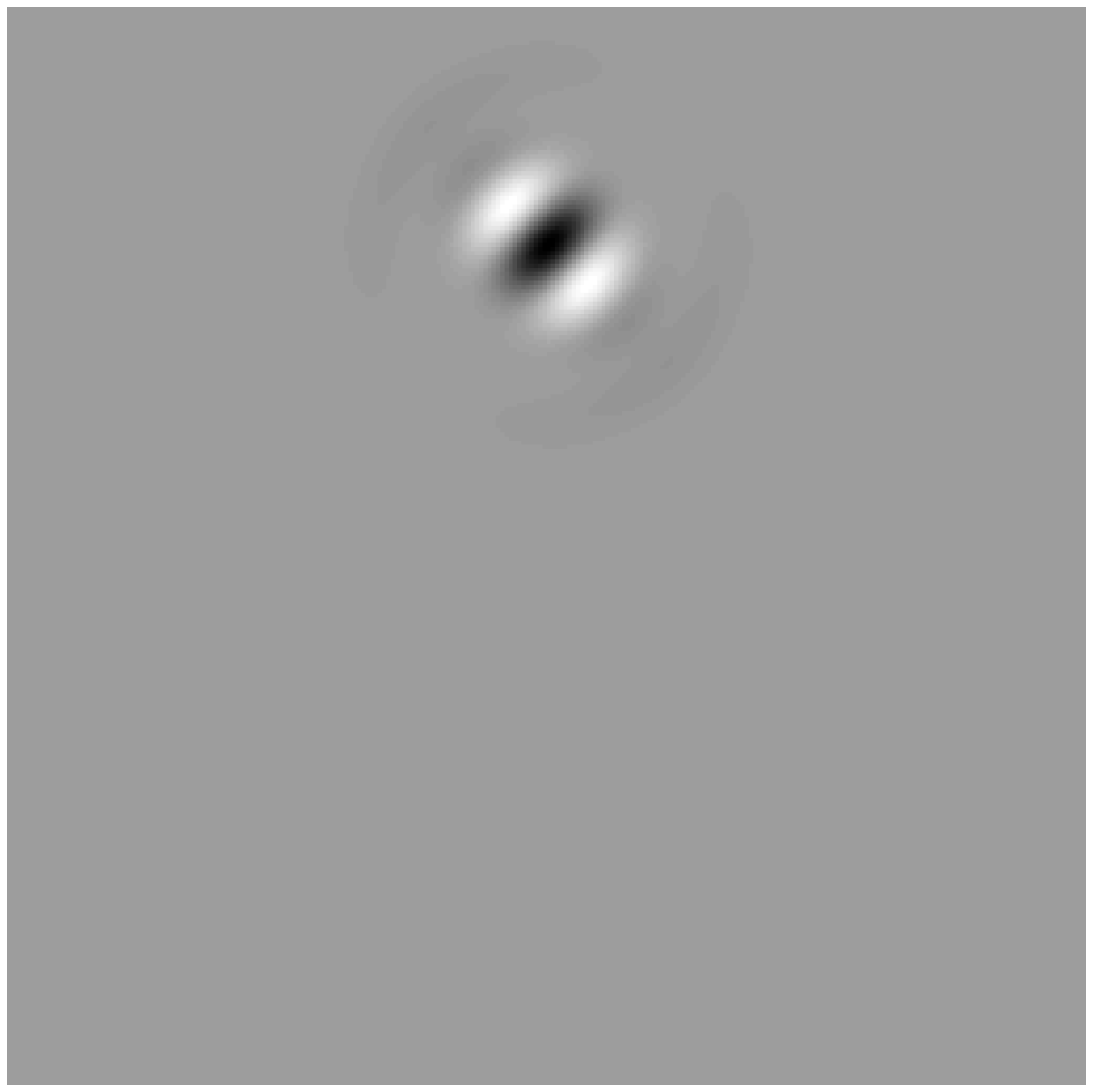}
\epsfxsize=2cm \epsfysize=2cm \epsfbox{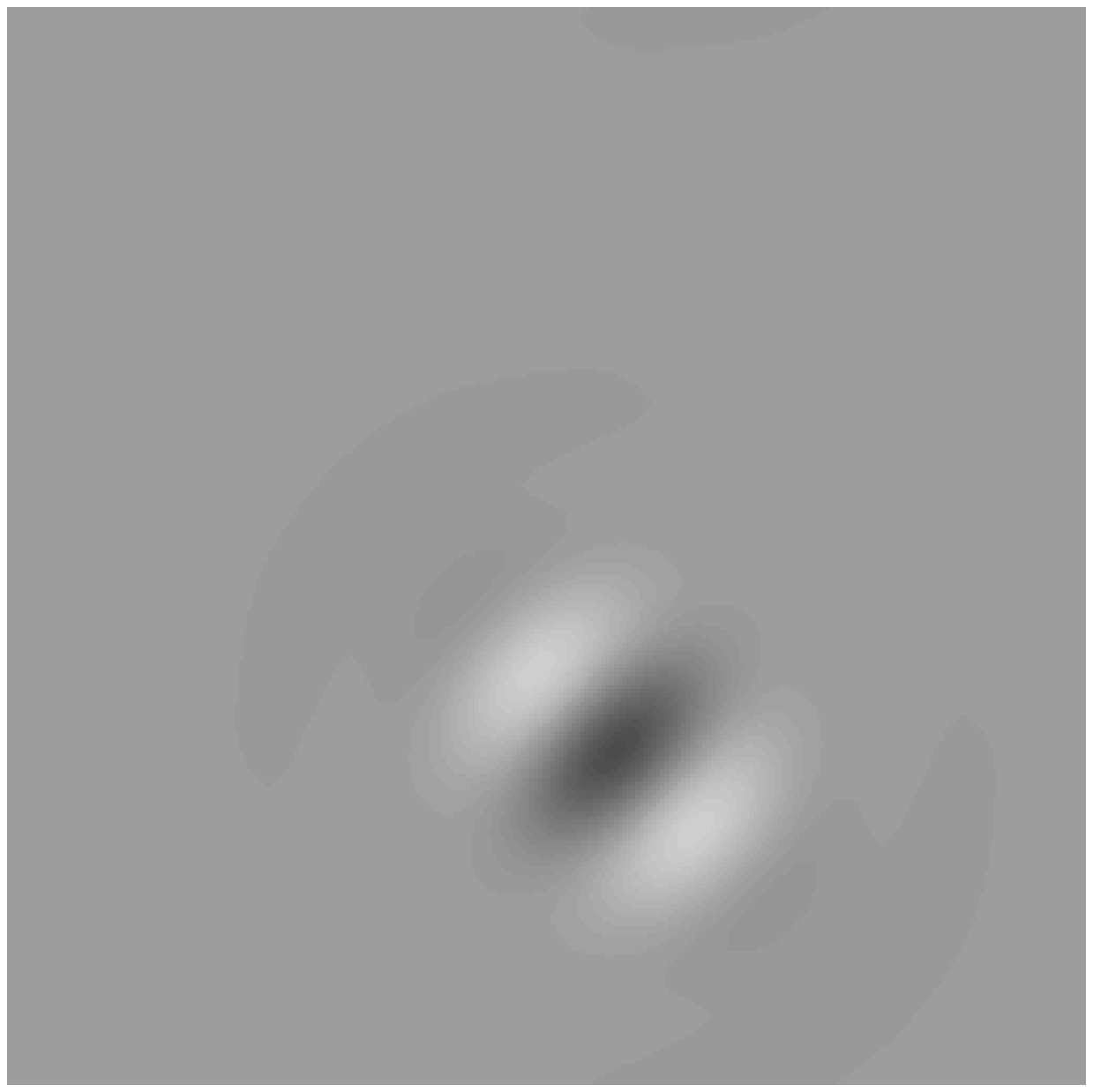}
\epsfxsize=2cm \epsfysize=2cm \epsfbox{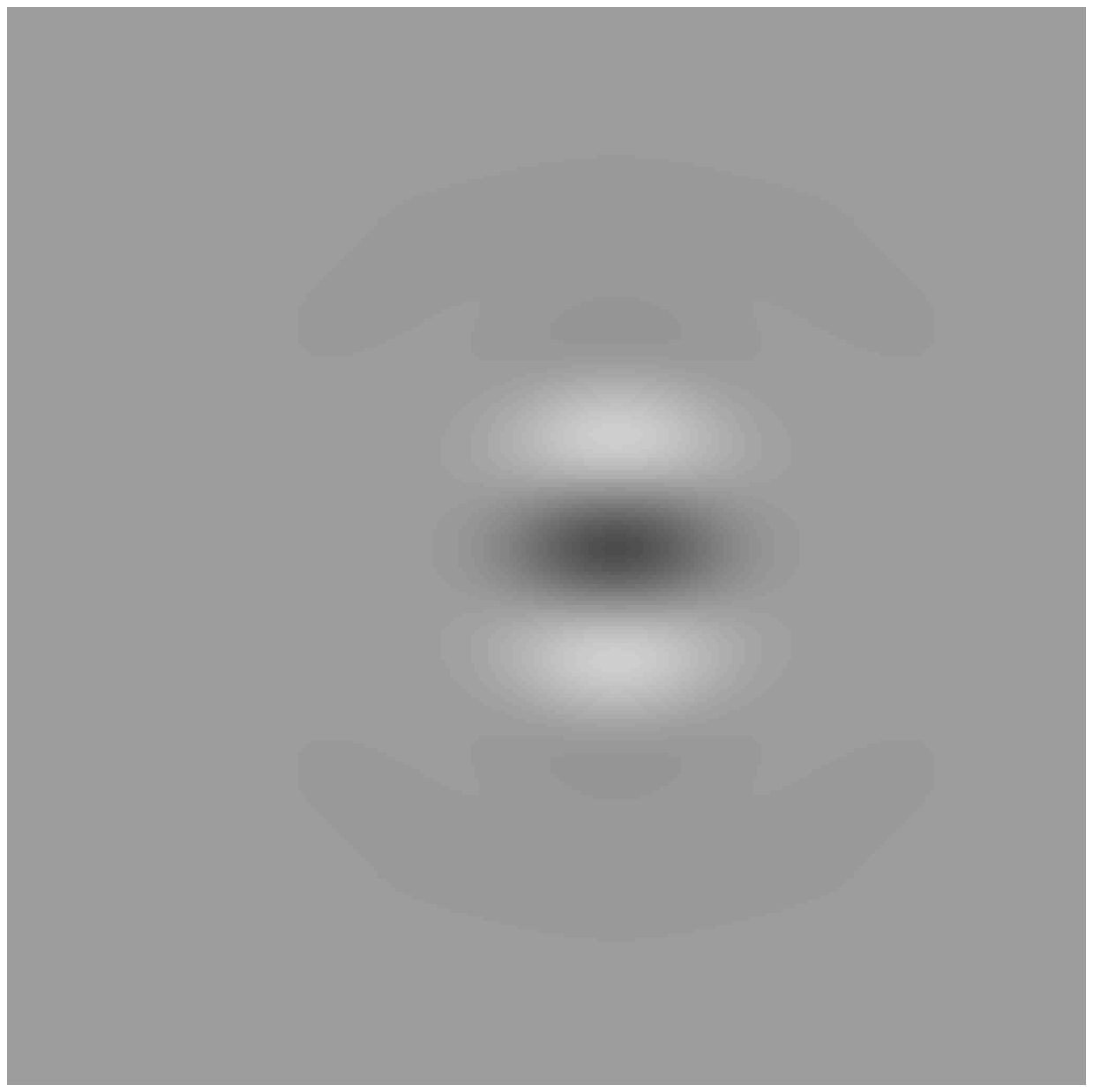}
\epsfxsize=2cm \epsfysize=2cm \epsfbox{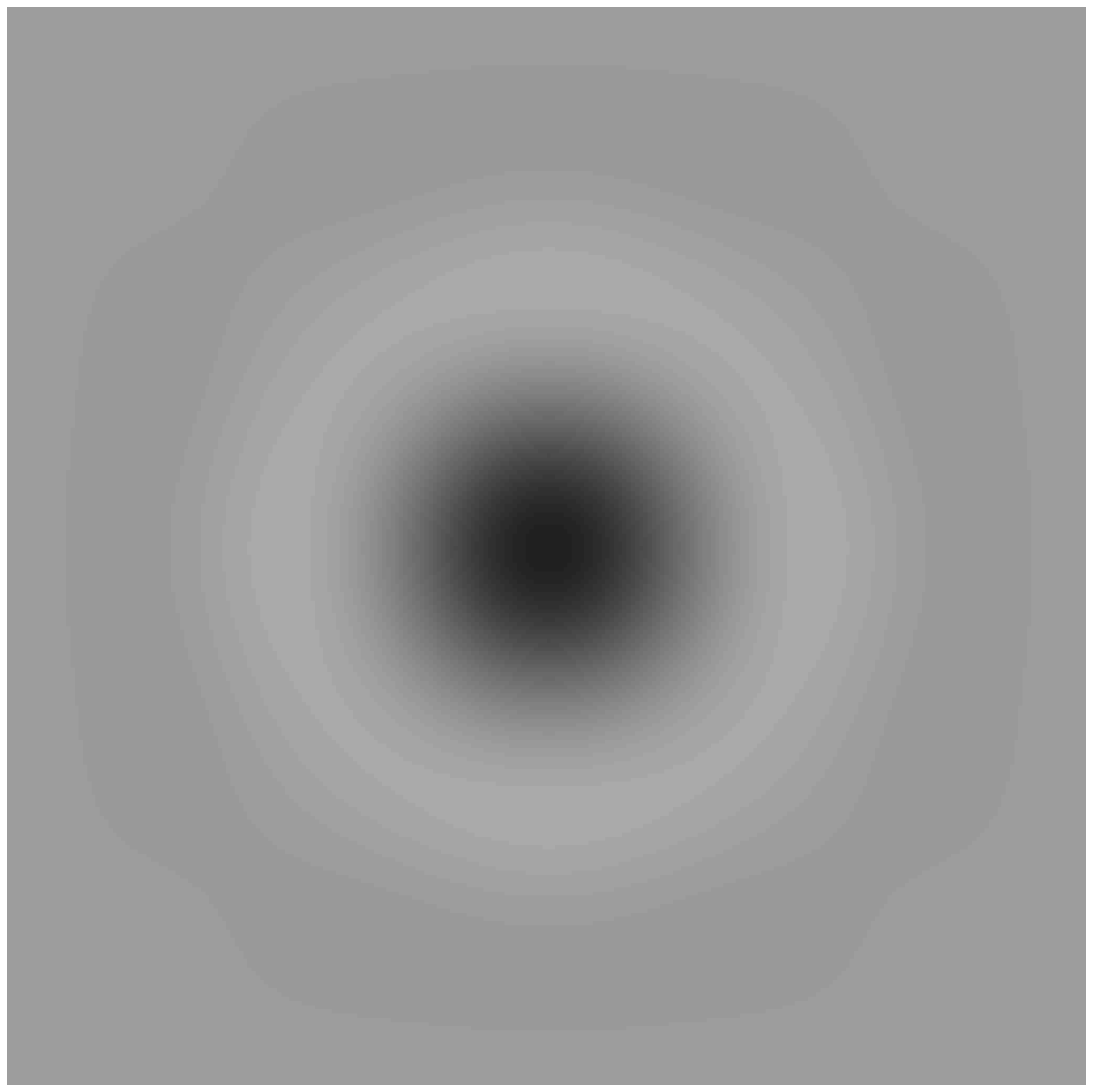}\\
\epsfxsize=2cm \epsfysize=2cm \epsfbox{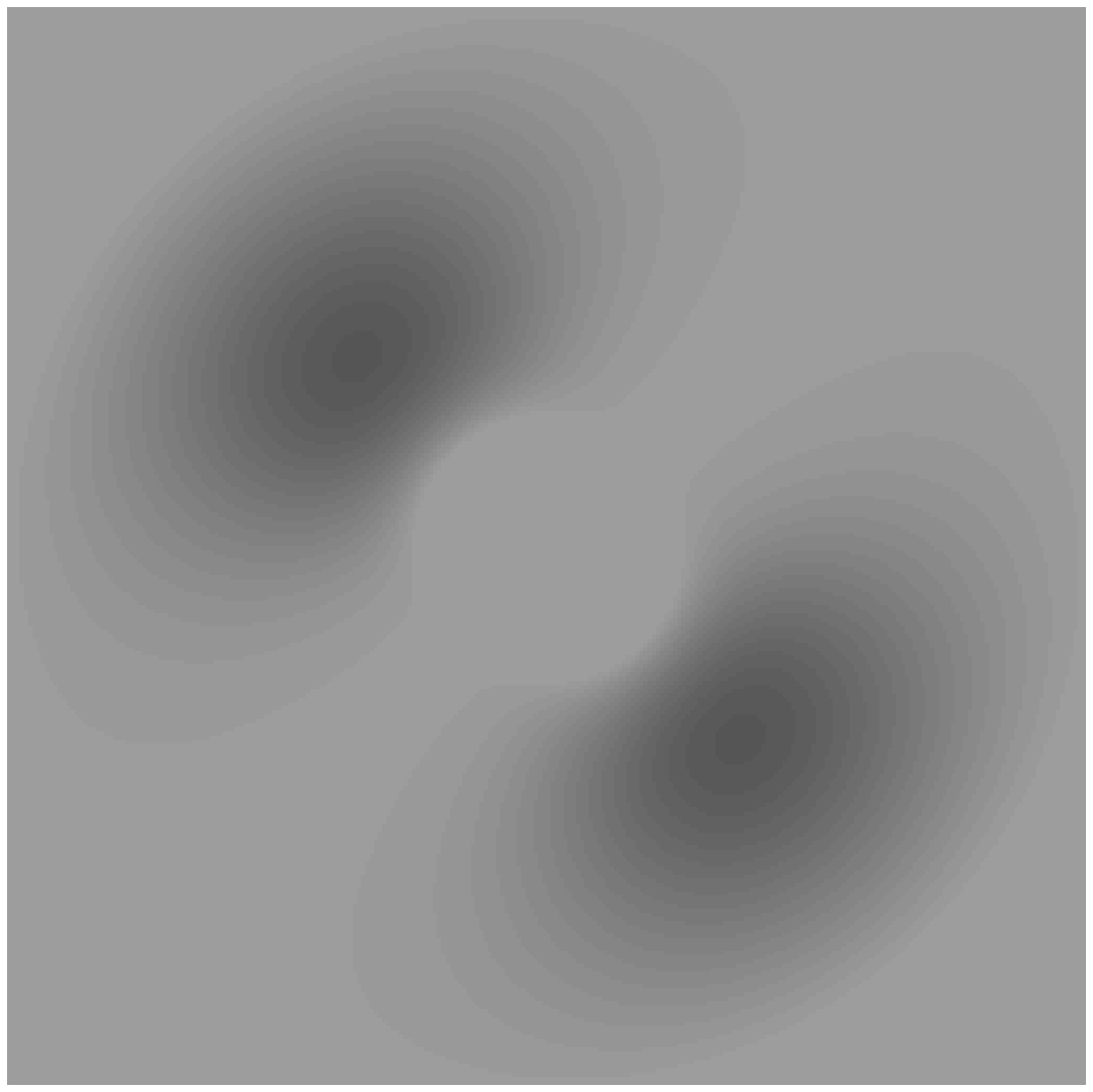}
\epsfxsize=2cm \epsfysize=2cm \epsfbox{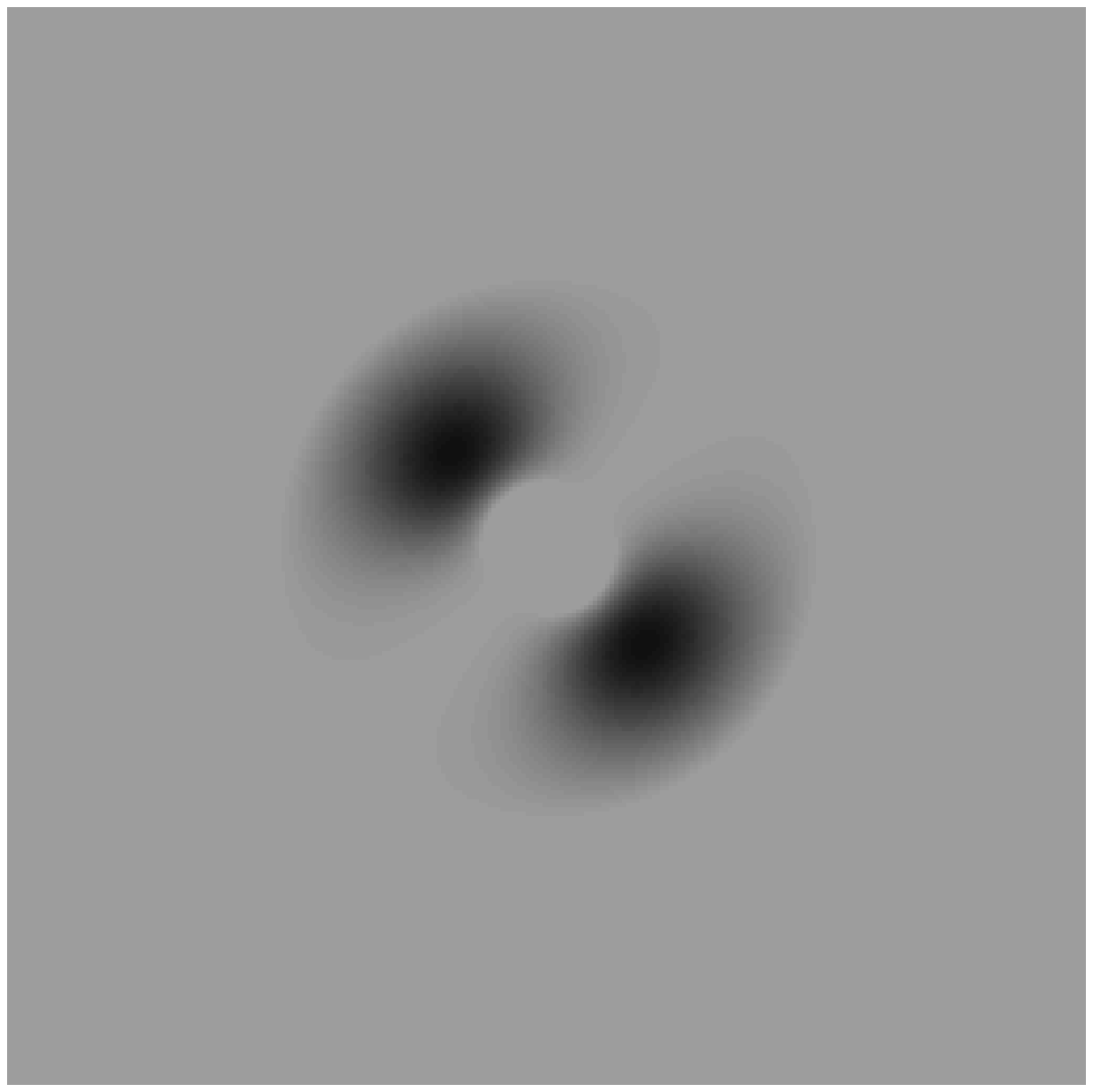}
\epsfxsize=2cm \epsfysize=2cm \epsfbox{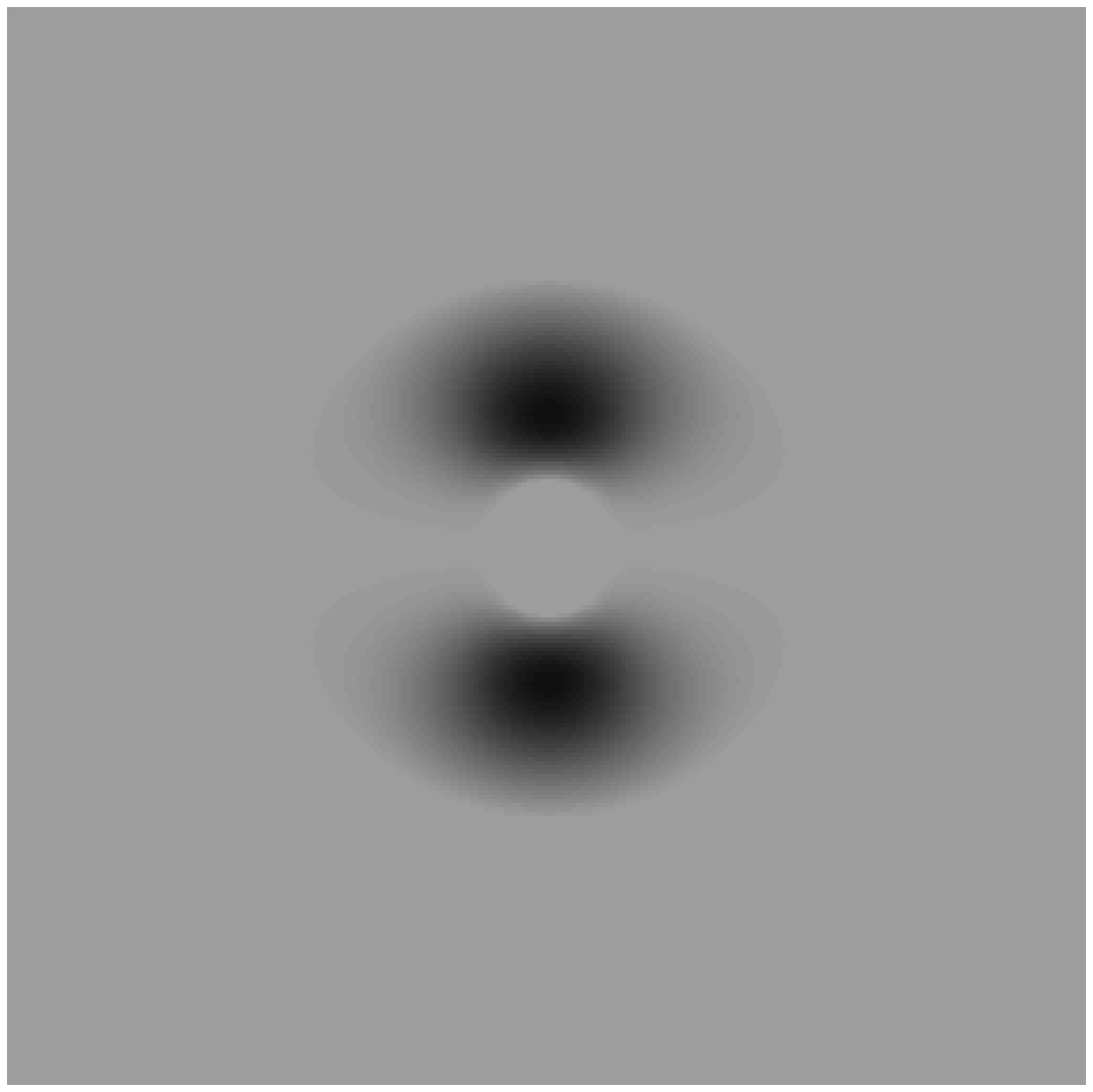}
\epsfxsize=2cm \epsfysize=2cm \epsfbox{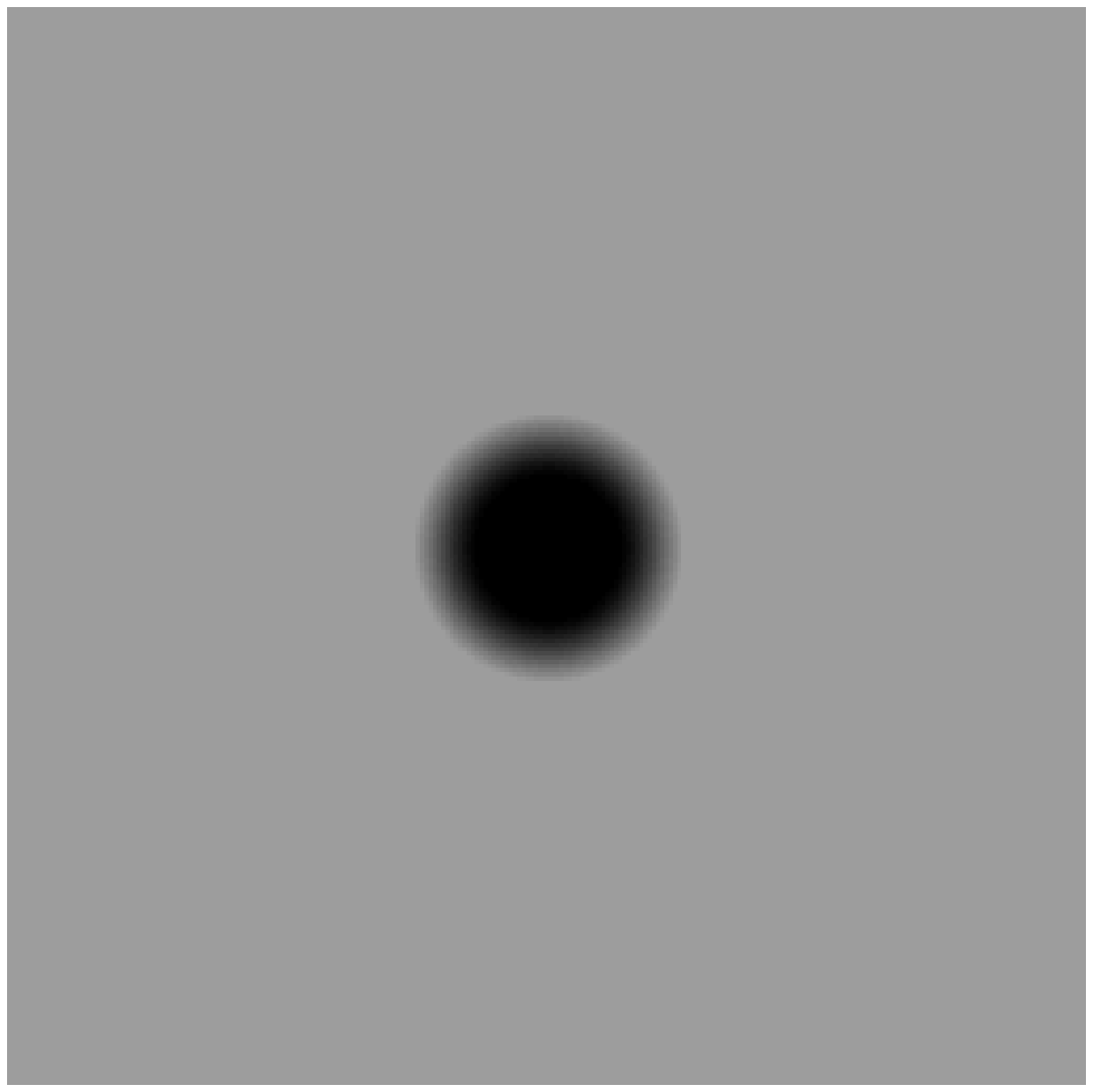}
\end{center}
\caption{Top row: wavelets in space; Bottom row: wavelets in Fourier plane. Left to right: wavelet at a fine scale $j+1$, centered at location $q_0$, oriented along the first diagonal; wavelet at a coarser scale $j$, centered at location $q_1$, oriented along the first diagonal; wavelet at the same coarser scale $j$, centered at location $q_2$, oriented along the horizontal axis; scaling function, centered at location $q_2$.}
\label{fig:wav}
\end{figure}

We use a two--dimensional overcomplete
 wavelet representation
 which is more adequate to the analysis of astrophysical images.
The wavelet decomposition of a signal $s$ in our case reads:
\be
s  = \sum_{q\in N^2}\langle s, \phi_{q}\rangle  \phi_{q} + \sum_{j=0}^{J} 
\sum_{m=1}^{M} \sum_{q \in 2^{-j} N^2} \langle s, \psi_{j,m,q}\rangle \psi_{j,m,q} 
\label{eq:wt}
\ee
where $\phi_{q}$ are the scaling functions, $\psi_{j,m,q}$ are the wavelets and
$\langle ,\rangle$ are scalar products. The sum $\sum_{q}\langle s, \phi_{q}\rangle \phi_{q}$
 is the projection of $s$ on the coarsest scale, i.e. a low-pass version of $s$.
 Each scaling coefficient $\langle s, \phi_{q}\rangle$ contains information about the signal 
$s$ at the coarsest scale and at a specific location in space $q$.
 For $j$ fixed, the sum $\sum_{m} \sum_{q} \langle s, \psi_{j,m,q}\rangle \psi_{j,m,q}$
 is the projection of $s$ on the scale $j$, i.e. a band-pass version of $s$.
 Each wavelet coefficient $\langle s, \psi_{j,m,q}\rangle$ contains information about the 
signal $s$ at the specific scale $j$, orientation $m$ and location in space 
$q$. As usual with wavelet transforms, changing scale is done by dilating, and changing 
location is done by translating the wavelet:
$
\psi_{j+1,m,q}(x)= \psi_{j,m,q}(2x) $ and $ \psi_{j,m,q}(x)= \psi_{j,m,0}(x-q)
$
Hence, scale $j+1$ corresponds to a frequency band that is twice as wide and for which the 
central frequencies are twice as large as that of scale $j$. On the other hand, in space, 
the wavelets at scale $j+1$ are better localized than at scale $j$ since they are more narrowly
 concentrated around their center $q$ (see Fig.\ref{fig:wav}, column 1 and 2). 
Unlike the 2-D Daubechies wavelets, the wavelets (and scaling function) we use here are defined
 in the Fourier plane. This ensures that they are well concentrated in frequency. Moreover, it 
enables us to introduce orientation by rotating the Fourier transform of the wavelet (see Fig.\ref{fig:wav}, column 2 and 3). If $\tilde {f}$ is the Fourier transform of $f$ and $(r,\theta)$ are polar coordinates, then:
$
\widetilde{\psi_{j,m,q}}(r,\theta)= \widetilde{\psi_{j,0,q}} (r,\theta-\frac{m\pi}{M})
$
The transform is therefore close to rotation invariant and computation is fast via FFT.
The Fourier transform of the wavelets and scaling functions read:
\bea
L(r) &=& \cos{({\pi \over 2} \log_2(r))}\delta_{1<r<2}+\delta_{r<1}  \ \ \hbox{ \scriptsize\it low pass}\\
H(r) &=& \sin{({\pi \over 2} \log_2(r))}\delta_{1<r<2}+\delta_{r>2} \ \ \hbox{ \scriptsize\it  high pass}\\
G_{M}(\theta) &=& {(M-1)! \over \sqrt{M[2(M-1)]!}}\left|2 \cos \theta\right|^{M-1} \ \ \hbox{ \scriptsize\it  oriented}\\
\widetilde{\phi_{0}}(r,\theta)&=& L(2r,\theta)\\
\widetilde{\psi_{j,m,0}}(r,\theta)&=& L({r \over {2^j}}) H({{2 r} \over {2^j}})G_M (\theta - \frac{m\pi}{M}) \ \  \ \ \ ^{j\geq0,}_{0\leq m<M}
\eea 
The set of all wavelets and scaling functions determines a redundant system (they are linearly dependent),
 however, the Plancherel equation holds.

\subsection{The estimator}
Formally, our goal is to estimate several processes (CMB, SZ, point sources) from their contributions in observations at different frequencies. We estimate the processes $\{x(p,\nu_0)\}_p$ from the observations $\{y(\nu)\}_\nu$ given that:
$
y(\nu)= \sum_p{f(p,\nu)x(p,\nu_0)*B(\nu)+N(\nu)}
\label{eq:obs}
$
where $x(p,\nu_0)$ is the template of the pth process
 at a given frequency $\nu_0$,
$f(p,\nu)$ is the frequency dependence of the pth process, 
$B(\nu)$ is the beam and $ N(\nu)$ is the frequency dependent white noise.
Our estimation method will rely on two principles to discriminate the contributions from different processes. The first one is that we know some statistical properties of the processes (e.g. the CMB and noise are Gaussian processes, while the clusters are not). The second one is that some spatial properties of the processes can be captured by modeling the coherence of wavelet coefficients.
 For example, clusters can be described as spatially localized structures with a high intensity peak. 
 To estimate a particular wavelet coefficient $x_{j,m,q}$, one describes the statistics of a neighborhood of coefficients around it by a Gaussian scale mixture. For example, 
$
{\bf x}_{j,m,q} =(x_{j,m,q},x_{j,m,q+1},x_{j,m,q-1}, x_{j-1,m,q})
$
is a neighborhood of coefficients around $x_{j,m,q}$. It contains wavelet coefficients at the same scale with close location, and at a close scale with the same location. The Gaussian scale mixture is the model:
\be
{\bf x} \equiv  \sqrt z {\bf u}
\label{eq:mod}
\ee
where ${\bf u}$ is a centered Gaussian vector of the same covariance as ${\bf x}$, the multiplier $z$ is a scalar random variable and the equality holds in distribution. ${\bf u}$ and $z$ are independent and $E\{z\}=1$. The covariance of ${\bf x}$ captures the spatial coherence of the process. The (non-)Gaussianity of the signal is captured by the distribution of the multiplier $z$.
To illustrate the idea for the reconstruction process, 
let us consider the simple case where we observe one process polluted by noise: $y  =  x +  N$ (in wavelet space). 
 ${\bf x}$ is a Gaussian mixture 
$ {\bf x} \equiv  \sqrt z {\bf u}$, with ${\bf u}$ a Gaussian vector. 
If $z$ was a constant, $z=z_0$, then $E\{{\bf x}|{\bf y}, z=z_0\}$, the Bayes Least square estimate of ${\bf x}_{j,m,q}$ given the observed vector ${\bf y}_{j,m,q}$ and $z$, would be the Wiener filter: 
\be
E\{{\bf x}|{\bf y}, z=z_0\} = z_0 {\bf C_x} (z_0{\bf C_x} + {\bf C_N})^{-1} {\bf y}
\ee
where ${\bf C_v}$ is the covariance matrix of the vector $v$. However in our model, $z$ is not a constant, so $E\{{\bf x}|{\bf y}\}$, the Bayes Least square estimate of ${\bf x}_{j,m,q}$, is a weighted average of the Wiener filters above:
\be
E\{{\bf x}|{\bf y}\} = \int_0^\infty{ p(z=z_0|{\bf y}) E\{\bf x} |{\bf y}, z=z_0 \}~dz_0
\label{eq:inte}
\ee
The weights are determined by the probability of $z$ given the observation ${\bf y}_{j,m,q}$, noted $p(z=z_0|{\bf y}_{j,m,q})$, which is computed via Bayes rule:
\be
p(z=z_0|{\bf y})=\frac{p({\bf y}|z=z_0) p_z(z_0)}{\int p({\bf y}|z=z') p_z(z') dz'}
\label{eq:pzy}
\ee
where $ p({\bf y}|z=z')$ is a centered Gaussian vector of covariance $z' {\bf C_x} + {\bf C_N}$, and $p_z$ is the probability distribution of $z$.
Following this procedure, one gets an estimate $E\{{\bf x}_{j,m,q}|{\bf y}_{j,m,q}\}$ for each neighborhood of coefficients ${\bf x}_{j,m,q}$. From this estimated vector, we only keep the estimate of central coefficient $x_{j,m,q}$.

In the case where $p(z)=\delta_D(z-0)$, the Gaussian scale mixture described in eq. (\ref{eq:mod}) reduces to a Gaussian process, which is an accurate model for the CMB signal.
Other signals, in particular the cluster signal, are typically non--Gaussian. In order to model them, we will need a more elaborate distribution for $z$. 
In this paper we use a distribution $p(z)$ that we derived from the input 
SZ maps with the technique described in Pierpaoli et al. (2004). 
The cluster's distribution $p(z)$ has a tail for high $z$ values 
which is caused by the high intensity points in the cluster centers.
By using  this distribution instead of the delta function (which would correspond to a Gaussian process) we are suggesting to the reconstruction method that
in the map there should be more ``high intensity'' points than in the 
corresponding Gaussian case with the same variance.
In Pierpaoli et al. (2004) we also describe other choices for $p(z)$ and conclude that
the final performance in reconstructing the cluster center doesn't depend on the specific shape of the distribution $p(z)$ provided that $p(z)$  has enough power in the ``high--intensity'' tail.

\section{Results} \label{par:res}
\begin{figure}
\begin{center}
\leavevmode
\epsfxsize= 7.5cm  \epsfbox{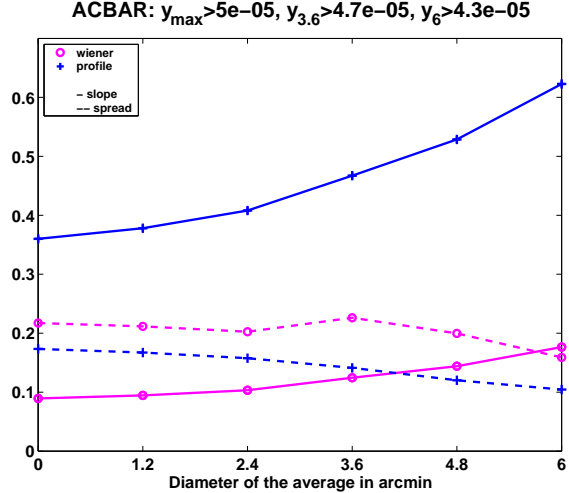}
\end{center}
\caption{ The  integrated output $y$ parameter versus the input one for the 
(12) largest clusters in the simulations. 
These values are obtained by using the ACBAR specifications.
The title  indicates the $y$ values of the less intense cluster considered:
$y_{max}$, $y_{3.6}$ and $y_{6}$ are the central intensities in  the input
 map with no smoothing  (the pixel size of the simulation being 1.17 arcmin)
and  with a smootihng angle of 3.6 and 6 arcmin respectively.  }
\label{fig:12}
\end{figure}

\begin{figure}
\begin{center}
\leavevmode
\epsfxsize= 7.5cm  \epsfbox{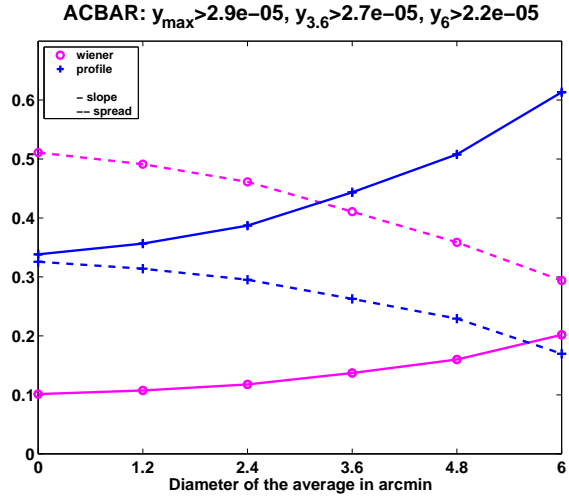}
\end{center}
\caption{ 
The  integrated output $y$ parameter versus the input one for the 
(34) most intense  clusters in the reconstructed maps. 
The values at the top have the same meaning as in fig.~2.
These values are obtained by using the ACBAR specifications. }
\label{fig:34}
\end{figure}

We applied the method described above to ACBAR simulated maps
with realistic beam and noise properties (see table \ref{tab:ic}).
We ignored  here the complications in the noise correlation matrix, and 
assumed instead white noise. We don't expect this approximation to 
highly compromise  the overall performance.
 In figures \ref{fig:12} and \ref{fig:34}
we show the performances of our method in  reconstructing the $y$ parameter 
of the largest and most intense clusters in the simulations.
The reason for such distinction is the following:
the method proposed here make use of the spacial covariances of the cluster's signal. 
Therefore  if the beam size of the experiment
is comparable to the typical cluster size (as is the case with ACBAR)
clusters with similar intensities but different core size may not be reconstructed equally well, since compact clusters are more likely to be confused with 
noise. The actual performance in the reconstruction largely depends on the noise level. For this reason we perform the two analysis for ``extended'' and
``intense'' clusters  separately.

We adopt the following procedure: first
we compute the input and output $y$ parameter integrated over a given angle 
(specified on the $x$ axis);
we then fit a line to the input/output values and calculate the average
departure of the output values from that line.
The figures report the slope of the fitted line (solid lines) and the scatter
about it (dashed lines) for the $p(z)$ cluster distribution (blue) and 
for a delta function corresponding to an hypothetical Gaussian signal 
(pink).

In these figures we notice that the Wiener filtering (which assumes the delta function for the $p(z)$ distribution) underestimates the 
high intensity peaks corresponding to massive clusters.
The estimator which uses the true  $p(z)$ distribution  performs on average 
 a factor of three better 
in reconstructing the central intensity of the clusters, 
quite independently from the integration  angle assumd.
The error average departure of the output value from the fitted line
is also reduced by about 30 per cent.
We conclude that the inclusion of non--Gaussian information greatly improves
the reconstruction of clusters SZ signal from observed maps.
Performances, however, depends on the specific characteristics of the experiment in hand, as well as on the characteristics of the clusters that we aim to reconstruct (i.e. total intensity and dimensions).
In the specific ACBAR case, by comparing figs~\ref{fig:12} and \ref{fig:34} we notice that extended clusters are on average slightly better reconstructed  and present 
a significantly  lower scatter. 

\section*{Acknowledgments}

EP is an NSF-ADVANCE fellow, also supported by NASA grant NAG5-11489.



\begin{thebibliography}{1}
\expandafter\ifx\csname url\endcsname\relax
  \def\url#1{\texttt{#1}}\fi
\expandafter\ifx\csname urlprefix\endcsname\relax\def\urlprefix{URL }\fi

\bibitem{SZ80}
R.~A. {Sunyaev}, I.~B. {Zeldovich}, {Microwave background radiation as a probe
  of the contemporary structure and history of the universe}, ARAA 18 (1980)
  537--560.

\bibitem{Herranz02}
D.~{Herranz}, J.~L. {Sanz}, M.~P. {Hobson}, R.~B. {Barreiro}, J.~M. {Diego},
  E.~{Mart{\'{\i}}nez-Gonz{\' a}lez}, A.~N. {Lasenby}, {Filtering techniques
  for the detection of Sunyaev-Zel'dovich clusters  in
  multifrequency maps}, MNRAS 336 (2002) 1057--1068.



\bibitem[{{Pierpaoli} {et~al.}(2004)}]{Pierpa04}
{Pierpaoli} E., {Anthoine} S., {Huffenberger} K., {Daubechies} I., 2004, 
submitted to MNRAS


\bibitem{Stoly02}
V.~{Stolyarov}, M.~P. {Hobson}, M.~A.~J. {Ashdown}, A.~N. {Lasenby}, {All-sky
  component separation for the Planck mission}, ApJ 336 (2002) 97--111.


\end{thebibliography}

\end{document}